# ACR: A CLUSTER-BASED ROUTING PROTOCOL FOR VANET

Saeid Pourroostaei Ardakani

Assistant Professor in Computer Science, Allameh Tabataba'i University, Tehran, Iran

## ABSTRACT

*Clustering is a technique used in network routing to enhance the performance and conserve the network resources. This paper presents a cluster-based routing protocol for VANET utilizing a new addressing scheme in which each node gets an address according to its mobility pattern. Hamming distance technique is used then to partition the network in an address-centric manner. The simulation results show that this protocol enhances routing reachability, whereas reduces routing end-to-end delay and traffic received comparing with two benchmarks namely AODV and DSDV.*

## 1. INTRODUCTION

Vehicular Ad-hoc Network (VANET) comprises of vehicular nodes connected via wireless links in an ad-hoc form. This is mainly deployed to provide transport and communication services. The key characteristics of VANET are outlined as below [1], [2], [3]:

1. High mobility: the nodes move frequently and with relatively high speed.

2. Patterned mobility: the nodes normally move through pre-defined paths e.g. roads and streets.

3. No energy constraints: VANET nodes are usually connected to continuous energy suppliers such as vehicle engine.

4. Frequent network topology change: node movements cause repeatedly topology changes due to high-rate connection link changes.

5. Unbounded network size: VANET is typically deployed large, dense and unbounded.

6. Frequent communications: the nodes repeatedly transmit network traffic to deal with network topology changes, routing and communications.

7. Time-critical for data delivery: the network traffic needs to be delivered within the time limit as VANET nodes need to perform actions or make decisions in real-time.

Each VANET node is typically attached to an On-Board communication Unit (OBU) which stays in the duty of providing communication with other nodes and/or Road-Side communication Unites (RSUs) [4]. RSUs allow communications in infrastructure level. Indeed, there are three types of communications in VANETS: (1) OBU-OBU (vehicle to vehicle), (2) OBU-RSU (vehicle to infrastructure) and (3) RSU-RSU (infrastructure level communications) [5]. The first, allows the vehicles to locally communicate in an ad-hoc manner and forward network traffic to each other. The second provides communication between the vehicles and RSUs to report network data such as lane traffic, road safety and/or congestion status. The third focuses on the communications between RSUs to transmit regional network data and provide globalization.

The Routing process includes route discovering, establishing and maintenance. This consists of two overlapping components which work in parallel: parallel mechanism and routing matrix [6].The former roots in the data forwarding procedure, the structure of control packets, network traffic transmission scheme, routing data storage mechanism, path discovery, detecting link failures and repairing the broken connections. The latter, works on the top of the former and is on the duty of best route selection when multiple paths are available.

VANET Routing focuses on establishing communication links between the OBUs and/or RSUs to transmit network packets [7]. The characteristics of VANET make routing different in this type of network with other ones such as Mobile Ad-hoc Networks (MANET) [8]: 1) VANET nodes consume energy with no concern, whereas MANET nodes typically need to consider energy consumption and utilize energy conservation techniques. 2) VANET nodes are highly mobile and typically move with relatively high speed, whereas MANET nodes are typically less mobile and move with limited speed. Due to this, network topology changes more frequently in VANET comparing with MANET. Hence, routing in VANET network needs to consider a number of additional parameters such as multi-path discovery and dynamic route establishment. 3) VANET mobility patterns are different with MANET as vehicular nodes usually move through a set of streets and highways which are predictable. In fact, VANET routing focuses only on the road maps which are different with MANET for which the nodes freely move.

In the remainder of this article, Section 2 explains and classifies VANET routing protocols to highlight their advantages, features and techniques. Section 3 presents ACR protocol and the key techniques which are used to resolve the existing drawbacks of VANET routing. Section 4 focuses on the experimental plans to test the performance of ACR. Section 5 evaluates the performance of ACR according to three metrics: routing reachability, traffic received and end-to-end delay. The results are measured and compared to two benchmarks namely AODV [9] and DSDV [10]. These protocols are selected as they are well-known in the literature, implemented and/or simulated a real world and adaptable for our experimental tools (OMNET++ [11]). This guarantees the correctness of the evaluation because ACR is compared against two routing protocols which are approved by OMNET++. Section 6 concludes the key advantages and disadvantages of ACR and then highlights the research issues which need to be addressed as future works.

## 2. LITERATURE REVIEW

VANET routing is the process of interconnecting the vehicular nodes directly or via logical routes (indirect) formed by intermediate nodes. Direct routing roots in the linking of two vehicular nodes which reside in the radio communication range of each other, whereas indirect communication is formed by a set of sequential and logical direct paths amongst the vehicular nodes. The routing protocols are classified into various categories according to different parameters such as network architecture, protocol proactively and operation [12]. Figure 1 shows the classifications of VANET routing. According to this, routing classification based on the protocol proactivity is the key as the other ones are parts of this. This classification comprises of three key categories: proactive, reactive and hybrid [13]. This section focuses on this classification of VANET routing.

## 2.1 Proactive Routing

This class of routing allows each node to continuously check and evaluate the potential paths to any other node in the network. Each node collects the routing information and keeps in its routing table. Proactive routing reduces routing delay as the required routing information to forward the network traffic can be immediately collected from the routing tables [14]. However, network resource consumption is increased as the routing information is continuously collected to establish new routes and/or update routing tables.

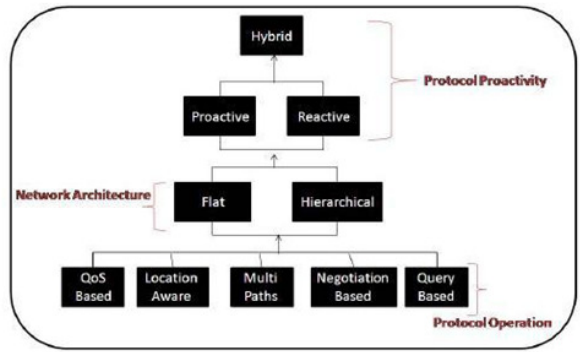

Figure 1: VANET Routing Classification

Destination-sequenced Distance-Vector (DSDV) [10] is an extended version of the original Bellman-Ford algorithm that proactively route network trafficusing the information stored in nodes' routing table. Each node maintains a routing table containing the destination nodes address, their distance in terms of hop count and a sequence number of route requests to find fresh routes and avoid loops. The routing tables are updated either in an event-driven or time-driven scheme. In the former, the routing data is collected when a particular event arises, whereas the latter gathers them at a particular time. Routing tables are updated in two forms: incremental and full-dump. Incremental updates only the changes/updated record of tables, whereas the whole routing table is updated in full-dump.

| Features | OSLR | DSDV |
|---|---|---|
| **Mechanism** | Link State | Distance Vector |
| **Architecture** | Flat | Flat |
| **Route Updates** | Periodical Hop-by-Hop | Periodical/Event-based Hop-by-Hop |
| **Communication** | Limited Broadcast (via MPSs) | Full Broadcast |
| **Key Advantage** | Reduced the route broadcasters | Fresh route |
| **Key Drawback** | Limited error recovery | High cost |
| **Routing Metric** | Shortest path | Shortest path |
| **Loop-free** | Yes | Yes |

Table 1: Proactive Routing Protocols

Optimized Link State Routing (OLSR) [15] is a link state table-driven routing algorithm that aims to decrease network traffic overhead by reducing the number of nodes that transmit the routing tables. Under this protocol, each node selects its single-hop neighbors, called Multipoint Relays (MPRs) to cover two-hop wireless communications. The route updates are firstly reported to MPRs. Then, MPRs propagate the updates throughout the network using User Datagram Protocol (UDP). OLSR greatly reduces the routing overhead comparing to flooding routing protocols as the number of nodes which transmit the route updates are decreased. OLSR is not very fault tolerance because UDP (as a communication medium) offers a limited error recovery service for communication. Table 1 compares and summarizes the key features of OLSR and DSDV.

## 2.2 Reactive Routing

Routing information is collected on demand if reactive protocols are used. A reactive path is established only when a node needs to forward network traffic through. This avoids to continuously consume the network resources to establish paths or update routing tables. Hence, resource consumption is usually reduced in reactive routing as compared to proactive. However, routing delay is increased in reactive routing as the network traffic needs to wait at each node until the required routing information is collected [14].

Ad-hoc On-demand Distance Vector (AODV) [9] establishes paths by forwarding routing messages from source to destination node in a unicast or multicast manner. This protocol focuses on maintaining the routing information of active links instead of available ones. Hence, the routing cost is reduced to a great extent. Each routing message contains the address of source node, a sequence number, the address of the destination node and the sequence number of the last routing message for the same path. Each node receiving the routing message replies the source node if a record of the requested path is found in its routing table. Otherwise, the route request is recorded in the node routing table and then the message is forwarded to the next hop node. The same procedure is performed at the next nodes until the destination is reached. The network traffic is forwarded from the source when the route reply message is received from the destination. Forwarding the routing information of only active paths decreases the routing overhead as compared to when the information of all available paths is forwarded. However, this reduces the fault tolerance as other potential or available route information is not recorded in the case of the active path failure.

Temporary Ordered Routing Algorithm (TORA) [16] establishes shortest paths between the source and distance node using distance values. First, TORA maps the network into a directed graph according to all the potential paths from the source to the destination node. Each graph link is allocated by a value which is called height. The height value is measured using a proportional proximity of the node to the destination. The route request is forwarded from the source node until it is received by either the destination node or any of its single-hop neighbors. Upon receiving the route request message, the destination node sends back a route reply message containing the height value to the source node. The value is increased by one at each intermediate node until the source node is reached. According to this, a greater value is allocated to the farther node, whereas the closer ones get a lower height value. TORA forwards the network traffic from the source to the destination node via the intermediate nodes whose height values are decrement. The key benefit of TORA is to deal with network topology changes. Any node detecting a link failure removes the path record from its routing table and then sets its own height to the highest value. The node broadcasts then a message to its single-hop neighbors informing them about the link failure. Any node receiving this message removes the record of the path from its table. However, this procedure increases memory usage and communication overhead in TORA comparing to AODV especially when the network deployed dense and network topology is changed frequently. Table 2 compares and summarizes AODV and TORA.

Table 2: Reactive Routing Protocols

| Features | AODV | TORA |
|---|---|---|
| Mechanism | Hop count/route sequence | Hop count |
| Architecture | Flat | Flat |
| Route Updates | As needed | Link/node failures |
| Communication | Uni/multicast | Local broadcast |
| Key Advantage | Reduced routing overhead | Reduced route updates |
| Key Drawback | Single route | High cost for large/dense networks |
| Routing Metric | Freshest/shortest path | Shortest path |
| Multiple Routes | No | Yes |

## 2.3 Hybrid Routing

This class of routing is the combination of proactive and reactive to minimize and maximize their weakness and strengths respectively. Proactive routing is suitable for the network with more stable connections, whereas reactive is appropriate for the network with high topology changes [17]. Hybrid routing combines the key aspects of proactive and reactive routing to enhance routing performance. However, it may inherit the existing drawbacks of both proactive and reactive routing that are high communication overhead and delay respectively.

Zone Routing Protocol (ZRP) [18] partitions the network into a set of particular zones and utilizes then two routing mechanisms to forward the network traffic through. The first mechanism is called Intra-zone Routing Protocol (IARP) which focuses on forwarding intra-zone packets, whereas the second one is the responsible for inter-zone communication and named Inter-zone Routing Protocol (IERP). A zone is formed around each node using an allowed hop value which is set by the network consumer in advance. Then, each node performs IARP to collect the routing information from its local zone. This mechanism is used to proactively route intra-zone packets. IERP is performed only by the nodes which reside on the border of each zone to inter-connect the zones. IERP provides reactive inter-zone communications. ZRP reduces the overhead of reactive routing as only a limited number of nodes (residing on the zones boarder) utilizes reactive routing instead of whole network. However, the allowed hop value is a key drawback of ZRP as this influences on the network resource consumption and routing delay. This means that the network resources are extremely consumed if the zone size is too large and IARP needs to collect a great deal of routing information, whereas the routing delay is increased if the zone size is set small and IERP collects routing information from a greater number of inter-zone links.

LANdMark Ad-hoc Routing (LANMAR) [19] proactively forms a set of zones in the network according to the nodes mobility patterns (i.e. direction and velocity) and then reactively forwards the network traffic through. Each zone is managed by a selected node which is calledthe landmark. Hence, each node maintains the information such as zone ID, landmark LD and the list of neighbors at its local routing table. The routing tables are updated if the network topology changes due to mobility or node failure. The source node sends a route request message containing the destination ID to the local neighbors. An intermediate node replies back the source node if the destination node is recognized. Otherwise, it forwards the route request message to its landmark asking for a connection to the destination. The landmark sends back the reply route message if the destination is found, otherwise the landmark sends the route request to other landmarks to find the destination. The availability of landmark nodes is a drawback of LANMAR as they should be available to reply back the route request messages. Besides, the routing delay and overhead is increased if the network deployed is dense and the topology changes frequently. This means that the landmark nodes need to repeatedly collect routing information from a number of neighbor nodes when the nodes are highly mobile. It results in increasing the routing delay and communication overhead. ZRP and MAYANMARare compared and summarized in Table 3.

Table 3: Hybrid Routing Protocols

| Features | ZRP | LANMAR |
|---|---|---|
| **Mechanism** | Hop count zone | Mobility pattern |
| **Architecture** | Flat | Hierarchical |
| **Route Updates** | Periodical | Event based or mobility |
| **Communication** | Limited Broadcast | Limited Broadcast |
| **Key Advantage** | Reduced messages and flexible | Mobility supported |
| **Key Drawback** | Allowed hop number | Landmark availability |
| **Routing Metric** | shortest path | Shortest/freshest path |
| **Loop-free** | Yes | Yes |

## 3. THE PROPOSED APPROACH

ACR presents a cluster-based routing protocol to forward network traffic over VANET. Under this protocol, each node firstly gets an ID, which is called LOCO, according to its location and mobility. Then, the network is partitioned into a set of clusters based on the nodes mobility pattern. Hamming distance technique is used to measure the similarities of nodes mobility using LOCO values. The nodes are grouped using a lightweight clustering algorithm if they have low hamming distance. Each cluster is managed by a Cluster-Head (CH) which stays in the duty of dealing with intra-cluster connections especially communicating with RSUs.

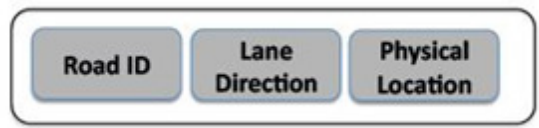

Figure 2: LOCO format

The key duty of ACR is to forward the network traffic mainly control and synchronization messages throughout a vehicular network. ACR utilizes two routing protocols: V2V and V2R. V2V utilizes a DSRC (Dedicated Short Range Communications) communication pattern to forward inter or intra-cluster traffic between the vehicular nodes. This allows the vehicle nodes communicate in an ad-hoc manner in which each node forwards the network traffic to any other ones residing in the same cluster. In fact, a network packet is dropped out if it is received by a node which resides in a different cluster. This reduces the network transmitted traffic and resource consumption (mainly energy and bandwidth) as unnecessary message broadcasts are estimated.

V2R supports back-end communications between vehicular nodes and the network infrastructure. By this, the CHs report the lane and road status to RSUs. This allows RSUs to inform back moving vehicles in the case of traffic jam or accident.

The value of LOCO is calculated using the node location, road ID and direction. This is generated at each node using the information taken from GPS and the local RSU. As Figure 2 shows, LOCO is represented in three fields: road ID, lane direction and physical location. Road ID is a binary address to show the road which the vehicle moves through. This is taken from the local RSU. Lane direction and physical location are received by the GPS. The former keeps the direction of vehicle, left-to-right or right-to-left for example, whereas the latter shows the location/region of the vehicle. LOCO is dynamically updated when the vehicle changes its location, road and direction.

## 3.1 Clustering Procedure

The network is partitioned using a distributed clustering algorithm in ACR. Clustering is a technique commonly used to establish hierarchical routing infrastructure in VANETs [20]. The key benefit of utilizing clustering technique in VANET is to reduce network traffic and congestion [21]. Clustering limits the network communications into a bounded area (clusters) which results in the reduction of message number. Furthermore, clustering technique allows the cluster-head to collect and aggregate packets that reduces the number of transmitted messages [22].This results in a reduction of transmitted network traffic, which decreases network resource consumption and congestion.

```
Data: Routing Table (RT), Hamming distance (HD), LOCO (X,
RID, L, PL)
/*X: sender ID, RID: Road ID, L: Lane
Direction, PL: Physical Location */

Clustering on Node Y:
LOCO info collected, a message broadcasted;
if LOCO received from X then
        Update RT (X, RID, L, PL, C);
        /*C column shows cluster-ship */
        HD.X ← HD(Y,X).RT(RID,L);
        /*Hamming distance value is
        measuredfor the receiver (Y) and
        sender (X) on RID and L */
        if(HD.X == 0) and (|PL.Y-PL.X| <= MAD) then
                Update RT(X, RID, L, PL, 'T');
                /*X and Y are now clustered */
        end
end
/* ************************************* */
Data: X.CH: cluster-head flag of node, RT.EoF: end of file)
CH Selection:
While(RT.EoF != null) do
        If(PL.Y |from PL.X) then
                Y.CH = 'F';
                break;
        end
        next RT.record;
end
```

Algorithm 1: ACR Clustering Procedure

According to Algorithm 1, each node firstly generates and broadcasts its LOCO to form the clusters in ACR. This allows a node to discover its vicinity using the received LOCO. The nodes which have similar LOCO reside in the close regions and have the potential to form the clusters. The similarity of LOCO is considered using Hamming distance. Hamming distance technique [23] is generally used to find the difference of binary values, by counting the number of flipped bits in fixed size binary data streams and returning the value of the difference as the distance. By this, the nodes which have no hamming distance (the value of hamming distance is zero) for road ID and lane direction have the potential to reside in the same region. Amongst all these nodes, the nodes whose physical distance is less than the maximum allowed radio range are partitioned as a cluster. This guarantees that all the cluster-members are connected in each cluster. Maximum Allowed Distance (MAD) is used to show the allowed physical distance between nodes in a cluster. MAD is defined by the consumer in advance of network deployment according to nodes' maximum allowed radio range. The maximum value of MAD is the radio range of vehicles' transceiver. Increasing MAD will subject to bigger size network cluster, whereas its reduction results in smaller size clusters.

To select cluster-heads (CHs), location information (LOCO) is used to find the nodes which move as the head of clusters. At each cluster, the node becomes CH if it moves in the front of all the cluster members. Due to this, each cluster member considers the received LOCO messages to find out the nodes which may move front. The node marks itself as CH if no one is found in front. The duty of CHs is to broadcast the road safety and traffic status to the cluster-members. They also report the situation of traversed road to the RSUs for further processing and utilization such as traffic information and accident report. RSUs are connected through telecommunication infrastructure and are in charge of transportation control. Each RSU is attached by an address which shows RSU's location and lane of coverage. According to Figure 3 RSU11 and RSU12 respectively cover left-to-right and right-to-left lanes of the road whereas RSU11.1 manages a sub-road of RSU11. RSUs utilize the information reported by CHs to control the road safety and update the vehicles which get connected next.

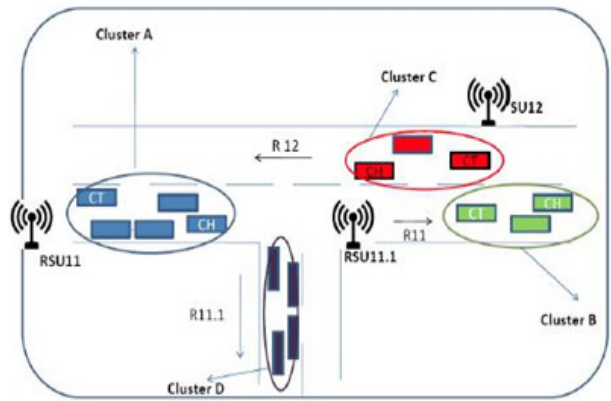

Figure 3: ACR Clustering Example

ACR performs particular mechanisms to deal with the network topology change and cluster reformation. The cluster topology may change due to CH leave, adding new vehicle and/or link failure. Unexpected network topology change mainly node failure due to hardware damage and/or node capture attacks are not addressed in this section, but they will be discussed as future work.

1. CH leaves: the leaving CH asks its neighbors to find the best fitted replacement if it wants to leave the cluster. The cluster members reply back the enquiry with their location information. The CH selects the closer one as the new CH and then sends a message to inform it. The new CH takes the role to communicate with RSUs and the cluster members.

2. New node joins: the new node broadcasts a request to join a cluster. It waits until a reply message (containing the location address of sender) is received from a clustered node. Then, the physical distance values of new node and the sender nodes are measured. The new node joins the cluster which the closest sender belongs to.

3. Link fails: the disconnected nodes may join other clusters or form a new one. A node is recognized as disconnected if it passes a RSU with receiving no update message from the CH. To deal with this, disconnected nodes firstly broadcast an enquiry to hear from the CHs. If a message is received, the disconnected nodes record the information and joins the cluster. The closest cluster is selected to join if messages are received from multiple CHs. Otherwise, the disconnected nodes form a new cluster using ACR clustering approach if no CH replies.

## 4. EXPERIMENTAL PLAN

To test and evaluate ARC, VEINS [24] framework is used. VEINS is an open source framework which work over two simulators: OMNET++ and SUMO. OMNET++ [11] is an open-source simulator for which there are implementations of AODV [9] and DSDV [10]. This uses modelling frameworks mainly MiXiM [25] and INET [26] to offer detailed models of radio wave propagation, interference estimation, radio transceiver power consumption and wireless MAC protocols. SUMO simulates road traffic for vehicular nodes.

The experiments utilize a random node distribution model to deploy the network over simulated roads. SUMO provides the road map in which the nodes freely move with random speed and lane direction. The nodes are allocated by a random speed value in three classes: 0- 4, 6-10 and 12-16 m/s. Each experiment runs for 200 times to get an acceptable level of confidence. The number of repetition is calculated using statistical power analysis technique [27]. This technique determines the necessary number of repetition using the standard deviation value (from a subset of samples/experiments which was 30 times with 5 nodes here) and according to a level of confidence (%90). Table4 presents details of the experimental plan.

| Element | Simulation Time | Repetition | Node count |
|---------|-----------------|------------|------------|
| Range | 2000s | 200 times | 100 |
| Element | Environmental Noise | Node distribution | Speed |
| Range | Enabled | Random | Slow(0-4)m/s Med(6-10) Fast(12-16) |
| Element | Communication range | Initialization time | Area size |
| Range | 100m | 100s | 1000X1000m |

Table 4: Simulation Setup Parameters

The experiments measure three metrics which are those typically used in the literature to evaluate the performance of routing protocols [14]: reachability, average end-to-end delay and total traffic received. The aim is to show how ACR improves upon AODV and DSDV in each of these aspects:

1. Reachability: this focuses on the number of reachable routes over all possible onesbetween source and destination nodes. Reachability is collected due to its high impact on the routing performance. This means that a routing protocol has a better performance if it has the potential to detect a greater number of reachable routes to forward the routing traffic. Hence, increasing reachability is addressed as a key objective in routing protocol design.

2. Average end-to-end delay (ETE): this calculates the average end-to-end delay of routing. ETE is measured from when the network packets leave the source nodes until they are collected at the target node. It is in influenced by communication delays such as packet reception/transmission and routing latency mainly route computation delay. ETE has the potential to influence the communication freshness. This means that the packets are expired or lose their usefulness if they are delivered late. For example, an accident arises if the break warning messages are slowly delivered to the rear vehicles.

3. Total traffic received: this represents the amount of network packets received in the entire network. The network messages are either control or data. Control packets are transmitted to deploy the network, establish/maintain the routing infrastructure and route the network packets including Hello, route request/reply, route error and maintenance, routing update and acknowledgement. Data packets are forwarded to transmit network information such as road status and break warning. Increasing the routing traffic results in higher network resource consumption. Furthermore, end-to-end delay (ETE) rises due to increased wireless channel access and communication delays when network traffic increases. Hence, reducing transmitted network traffic is required to reduce network resource consumption and end-to-end delay.

## 5. RESULTS

This section evaluates the performance of ACR, AODV [9] and DSDV [10] based on the routing performance metrics that are described in the previous section.

### 5.1 Reachability

Reachability is calculated as the proportion of successful route discovery attempts over the total number of route discovery operations. As Figure 4 illustrates, reachability changes due to increasing node speed. Reachability is correlated to the node speed as the route discovery success is influenced by the network topology change. This means that reachability reduces when the nodes velocity is increased. Increasing node speed results in increasing network topology changes. This reduces reachability especially if the routing information is not dynamically collected by the routing protocol.

Route reachability increases in reactive routing as compared to proactive. This is because of dynamic routing information collection according to the nodes mobility. This means that the chance of route discovery failures is reduced in reactive routing as the road is established on demand using the most up-to-date routing information. On the other hand, route reachability reduces in proactive routing protocols such as DSDV because the nodes routing table are not updated as quick as the network topology changes. In other words, the network topology changes frequently and the proactive routing cannot update the routing tables. Hence, the nodes may use out-of-date routing information which usually has low chance to reach the destination.

ACR outperforms AODV and DSDV in terms of reachability as it routes the network traffic over clustered infrastructure. ACR partitions the network according to the nodes mobility pattern. Each cluster can be considered as a supper-node which moves throughout the network. This reduces the impact of node mobility on the network topology. Under ACR, the node mobility does not change the network traffic unless it leaves its cluster. Hence, the number of routing table updates is reduced as the network topology changes only due to the change of cluster (super-node) mobility pattern and not a particular vehicle. This increases ACR reachability comparing to AODV and DSDV.

This increases ACR reachability comparing to AODV and DSDV.

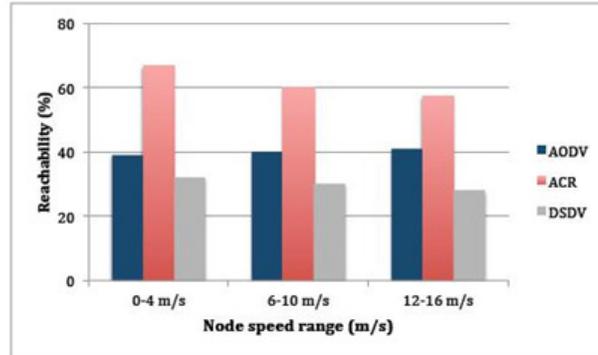

Figure 4: Reachability Ratio

## 5.2 Average End-to-End Delay

Average End-To-End delay (ETE) is measured as the average time from when a packet leaves the source node until it is received by the destination. The objective of routing protocols is to reduce the average delay as it enhances data freshness.

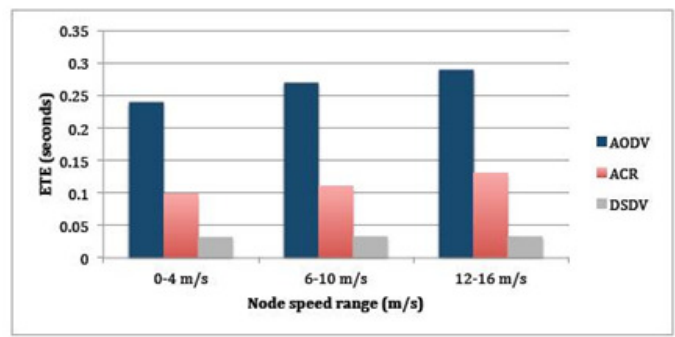

Figure 5: Average End-to-End delay

According to Figure 5, ACR outperforms AODV in terms of delay. This is because of utilizing clustering technique to partition the network and proactive routing information collection in each cluster. As the figure shows, DSDV reduces ETE delay comparing with AODV because of proactive routing. Proactive routing reduces ETE as the required routing information already is collected and intermediate nodes do not need to collect the information reactively during the routing.

ACR increases ETE as compared to DSDV because it uses reactive inter-cluster routing. This means that on-demand routing information over inter-cluster paths results in increasing ETE in ACR as compared to DSDV which uses proactive routing. Inter-cluster links are established if CHs are linked to forward data packets. In this case, CHs need to reactively collect routing information and this results in increasing ETE.

### 5.3 Total Routing Traffic Received

Routing traffic received has a high impact on the routing performance due to three key reasons:

1. Increasing the network traffic results in higher network resource consumption: both sender and receiver need to use the network resources such as bandwidth and wireless medium to transmit network packets. Hence, the network resources are increasingly used if the amount of routing traffic is increased.

2. Increasing routing traffic has the potential to reduce routing throughput: this increases network congestion and message failure if the nodes simultaneously access to the physical medium for communication.

3. Increasing routing traffic increases ETE: increasing network traffic increases waiting time to access the wireless channels and consequently increases ETE. The network packets need to be queued if the wireless channels are not available to transmit data. Hence, the waiting time (caused by idle-listening) to transmit data at the network nodes increases when the network traffic is increased. Furthermore, increasing routing traffic increases data packet failures and consequently packet delivery time. The probability of message failure is increased due to message collisions when routing traffic rises. Hence, nodes need to re-transmit routing packets until they are correctly delivered to the destination.

As Figure 6 shows, routing traffic received is increased in DSDV as compared to ACR. Clustering limits the network communications into bounded regions (clusters). A node receives network packets via intra-cluster link if it resides in the same cluster. This results in a reduction of transmitted network traffic comparing with at network which any node in the radio range of sender node receives the packets. In addition, transmitted network traffic decreases in clustered networks as intra-cluster messages are usually aggregated at CHs and the nodes do not need to individually transmit traffic. Due to the reasons, ACR reduces total traffic received comparing with DSDV.

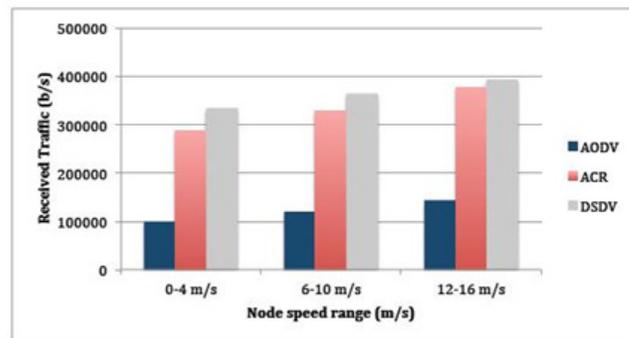

Figure 6: Total routing traffic received

ACR under-performs transmitted traffic as compared to AODV. This is because ACR utilizes proactive intra-cluster routing. This increases the number of network messages to update routing tables at the vehicles due to network topology change.

## 6. CONCLUSION AND FURTHER WORK

ACR partitions vehicular networks based on nodes mobility pattern. The objective of this protocol is to enhance routing performance and reduce communication overhead. Under this protocol, each node is allocated with an address according to its location and mobility. Then, a lightweight technique, called Hamming distance, is used in a distributed manner to form the cluster. Using clustering, this is not required that all nodes communicate with RSU to report road status. But, RSUs are only updated when CHs report. This would results in reduction of network traffic received comparing to DSDV which routes packets over at networks. ACR utilizes intra-cluster proactive routing to decrease ETE comparing to AODV which is a reactive routing protocol. Furthermore, ACR reduces the impact of node mobility on network topology change as this group's nodes based on mobility pattern. In fact, ACR forms a network of mobile clusters each of which independently move. Hence, the mobility of each node does not lead to network topology change, but a cluster mobility pattern makes. This increases reachability as compared to the routing table in which a single node may change network topology.

In future, the performance of ACR needs to be improved further when the speed of vehicle nodes increases. According to the results, increasing the nodes speed results in increasing the network traffic received and ETE. This increases the risk of ACR for real applications because these may lead to communication delay amongst the crowded vehicles.

Investigating the correlation between the cluster size and network performance can be addressed as a further work. Increasing the number of nodes in the clusters increases the proactive intra-cluster communications to establish the communication links and forward data packets. In other words, the routing overhead increases if the network deployed densely and the clusters are crowded. This results in the reduction of ACR performance. For this, the further experiments are required to figure out how well ACR performs if the network gets crowded.

Multi-level clustering may offer benefits to reduce communication overhead in ACR. Reactive inter-cluster routing is the key reason to increase ETE in ACR. In other words, routing latency is reduced if the number of reactive inter-cluster links is reduced in ACR. For this, utilizing a hierarchical clustering approach to form multi-level clusters may result in the reduction of ETE. In fact, combining the clusters which have similar mobility pattern and reside in close regions as super-clusters would reduce the number of clusters and consequently decreases reactive routing. However, this may subject to a trade-off with network resource consumption as proactive intra-cluster links increase.

## REFRENCES

**Authors:**

SaeidPourroostaeiArdakaniworks inAllamehTabataba'i University (ATU), Tehran, Iran, as an assistant professor in computer science. He joined ATU at 2015. Saeid's research is not limited to but mainly focuses on Ad-hoc networks, Routing, Internet of Things (IoT) and Cloud computing.Saeid received his PhD (focusing on Wireless Sensor Network Routing) in computer science from university of bath, UK, at 2015.

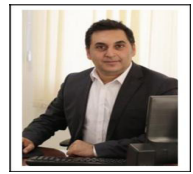